\documentstyle[epsf,twocolumn,prl,aps]{revtex}

\begin{document}
\draft

 \twocolumn[\hsize\textwidth\columnwidth\hsize\csname @twocolumnfalse\endcsname

\title{The Double-Time Green's Function Approach to
the Two-Dimensional Heisenberg Antiferromagnet with Broken Bonds} 
\author{ Yun Song$^{*}$, H. Q. Lin,
and Jue-Lian Shen$^{**}$}
\address{Department of Physics, 
The Chinese University of Hong Kong,
Shatin, N. T., Hong Kong, 
China}
\date{\today}
\maketitle
\begin{abstract}
\noindent

We improved the decoupling approximation of the  double-time 
Green's function theory, and applied it to study the 
spin-${1\over 2}$ two-dimensional antiferromagnetic Heisenberg
model with broken bonds at finite temperature. 
Our decoupling approximation is applicable to the spin systems
with spatial inhomogeneity, introduced by the local defects,
over the whole temperature region.
At low temperatures, we observed that the quantum fluctuation
is reduced in the neighborhood of broken bond,
which is in agreement with previous theoretical expectations.
At high temperatures our results showed that the quantum
fluctuation close to the broken bond is enhanced.  
For the two parallel broken bonds cases,
we found that there exists a repulsive interaction between the two parallel 
broken bonds at low temperatures. 

\end{abstract}

 ]

\section{Introduction}

In recent years, the two-dimensional (2D) antiferromagnetic (AF) 
spin system on a square lattice
has been one of the subjects of major interest in condensed matter physics
\cite{Manousakis}.
This follows from the experiments that copper oxide sheets in the
high-T$_{C}$ superconductors show strong AF correlations \cite{Kampf}. 
The undoped copper oxide materials are layered AF insulators and well 
described by the 2D AF Heisenberg model \cite{Manousakis}.
Doping holes into these materials leads to frustration of spins and 
ultimately to destruction of AF long-range order (LRO).
In the extreme limit of static hole, holes act as local 
defects and the inhomogeneous Heisenberg model is believed to 
describe some of the physics \cite{Aharony}. 

Several numerical and analytical works have been devoted to the effects of 
isolate defects, e.g., static vacancies \cite{Scalapino,Nagaosa,SDS}, 
broken or ferromagnetic bonds \cite{Schlottmann,Sandvik,Camley,Kotov}, 
and dynamic holes \cite{Dagotto}, on the magnetic properties of the 2D 
antiferromagnet.
The inhomogeneous Heisenberg systems are mainly divided into two types.
One is the site-defect (SD) model and the other one is the 
bond-defect (BD) 
model. These models are important for many fundamental problems, 
such as frustration, phase separation, and spin glass.
Here we adopt the BD model to study the effects of broken bonds
replacing AF links in the spin-${1\over 2}$ Heisenberg 
model at finite temperature.  
For zero temperature, Lee and Schlottmann have studied this model 
using the linear spin wave (LSW) theory \cite{Schlottmann}. 
They observed that the quantum
fluctuation is reduced in the neighborhood of the impurity link and the 
local magnetic moment is enhanced, in agreement with results obtained by Bulut 
$et$ $al$ for static vacancy case \cite{Scalapino}.

At any finite temperature, Mermin and Wagner \cite{Mermin} have proved that
for models such as the AF Heisenberg model the AF LRO is destroyed
by strong thermal fluctuations in low dimensional systems.
Spin wave theory which is based on the existence of LRO can not be 
directly used at finite temperature.
Alternatively, Kondo and Yamaji \cite{Kondo} proposed a
second-order Green's function (SOGF) theory to study the low dimensional  
Heisenberg model over the whole temperature region.
At high temperatures, this theory reproduces the correct results obtained 
by the high temperature expansion method \cite{Rushbrooke}. On the other hand, 
the results at low temperatures are similar to those of the 
modified spin-wave theory \cite{Takahashi}.
The SOGF theory does not violate the sum rule and rotation symmetry of 
spin correlations.
In the SOGF theory, the decoupling approximation is at a 
stage one step further than Tyablikov's random-phase approximation 
(RPA) \cite{Tyablikov}.
The SOGF theory has been successfully applied to various 
low dimensional homogeneous systems without LRO, 
such as the one-dimensional (1D) Heisenberg model \cite{Kondo},
the 1D XXZ model \cite{Zhang}, the 2D Heisenberg model \cite{Shimahara} 
and the 2D XXZ model \cite{Fukumoto}. Their results are in qualitative 
agreement with those numerical results \cite{Bonner,Okabe} over the 
whole temperature region. 

In this paper we extend the SOGF theory to study the inhomogeneous
Heisenberg model by improving the decoupling approximation.
The decoupling approximation proposed by Kondo and Yamaji (KYDA) \cite{Kondo} 
can not be directly applied to the inhomogeneous case without modification.
In order not to violate the sum rule in the inhomogeneous case,
we introduce an improved decoupling approximation, which is equivalent to
the KYDA in the homogeneous case. 
In our approximation two parameters $\beta_{i}$ and $\beta_{j}$ are 
attached to the correlation function of the two corresponding spins 
on sites $i$ and $j$ as 
$\beta _{i}\langle S_{i}^{+}S_{j}^{-}\rangle \beta_{j}$.
While in the KYDA, only one parameter $\alpha_i$ is introduced.
It is clear that these two parameters account for vertex correction
of spin-spin correlation and have to be introduced in 
order not to violate the sum rule of correlation functions. 
Thus, in our decoupling scheme, $N$ vertex correction parameters ($\beta_{i}$)
for a lattice of $N$ sites were introduced and no other extra parameter.
In the homogeneous case, we obtain $\beta ^{2}=\alpha$ and our 
decoupling approximation reduces to the KYDA.
Applying this method to the spin-${1\over 2}$ 2D antiferromagnetic 
Heisenberg model with broken bonds, we obtain reasonable results 
over the whole temperature region.  

The present paper is organized as follow: in Sec. II we present our extension
of the SOGF theory to the inhomogeneous spin systems and discuss the improved
decoupling scheme. 
Our numerical results
for some particular configurations of one, two, and three broken bond
cases are studied in Sec. III.
Finally, we conclude our findings in Sec. IV.

\section{Equation of Motion and Decoupling Approximation}

The 2D spin-$\frac{1}{2}$ Heisenberg model with bond-dependent exchange
constants can be expressed by the Hamiltonian
\begin{equation}
H=\sum_{\langle ij\rangle }J_{ij}\{\frac{1}{2}
(S_{i}^{+}S_{j}^{-}+S_{i}^{-}S_{j}^{+})+S_{i}^{z}S_{j}^{z}\},
\end{equation}
where $\langle ij\rangle $ denotes a sum over nearest-neighbor 
(NN) bonds,
and $J_{ij}$ is the exchange interaction between spins on sites $i$ 
and $j$. For the homogeneous case, $J_{ij}$ is equal to $J$ for all bonds.
For the inhomogeneous case we are studying here, $J_{ij}$ equals to zero
for broken bond, while it equals to $J$ for unbroken bonds.

We define spin Green's functions $G$ by
\begin{equation}
G(i-j,~t-t^{\prime })\equiv 
\langle \langle S_{i}^{z}(t);~S_{j}^{z}(t^{\prime })\rangle \rangle .
\end{equation}
After time-Fourier transformation,
the equation of motion of the spin Green's function $G$ can be evaluated
as 
\begin{equation}
\omega G(i-j,~\omega )=\sum_{\eta }J_{i,i+\eta }\langle \langle
S_{i}^{+}S_{i+\eta }^{-}-S_{i+\eta }^{+}S_{i}^{-};~S_{j}^{z}
\rangle\rangle,
\end{equation}
with $\eta $=\^{x}, \^{y}. The SOGF appears on the right hand side of 
Eq.(3). Furthermore, establishing equation of motion of the SOGF, we 
have
\begin{eqnarray}
\omega &\langle \langle &S_{i+\eta }^{+}S_{i}^{-}-S_{i+\eta
}^{+}S_{i}^{-};~S_{j}^{z}\rangle \rangle 
\nonumber \\
&=&
2\langle
S_{i}^{+}S_{i+\eta }^{-}\rangle (\delta _{i+\eta ,j}-\delta
_{i,j})
+2J_{i,i+\eta }\langle \langle S_{i}^{z}-S_{i+\eta
}^{z};~S_{j}^{z}\rangle \rangle   \nonumber \\
&&+\sum_{\eta ^{\prime }\neq \eta }\{2J_{i,i+\eta ^{^{\prime }}}[\langle
\langle (S_{i+\eta ^{\prime }}^{+}S_{i+\eta }^{-}+S_{i+\eta ^{\prime
}}^{-}S_{i+\eta }^{+})S_{i}^{z};~S_{j}^{z}\rangle \rangle 
\nonumber\\
&&\hspace{0.3in}-\langle \langle (S_{i}^{+}S_{i+\eta
}^{-}+S_{i}^{-}S_{i+\eta }^{+})S_{i+\eta ^{\prime }}^{z};~S_{j}^{z}\rangle
\rangle )] \nonumber\\
&&+2J_{i+\eta -\eta ^{\prime },i+\eta }[\langle \langle
(S_{i}^{+}S_{i+\eta }^{-}+S_{i}^{-}S_{i+\eta }^{+})S_{i+\eta -\eta ^{\prime
}}^{z};~S_{j}^{z}\rangle \rangle  \nonumber\\
&&\hspace{0.3in}+\langle \langle (S_{i}^{+}S_{i+\eta -\eta ^{\prime
}}^{-}-S_{i}^{-}S_{i+\eta -\eta ^{\prime }}^{+})S_{i+\eta
}^{z};~S_{j}^{z}\rangle \rangle ]\},
\nonumber \\
\end{eqnarray}
and the third-order Green's functions appear on the right-hand side of Eq. (4).

In the homogeneous case, Kondo and Yamaji decoupled the chain of 
equation (4) in an approximate way \cite{Kondo}, for example 
\begin{equation}
\langle \langle S_{m}^{+}S_{n}^{-}S_{i}^{z};~S_{j}^{z}\rangle \rangle
\rightarrow \alpha \langle S_{m}^{+}S_{n}^{-}\rangle \langle
\langle S_{i}^{z};~S_{j}^{z}\rangle \rangle   
\end{equation}
for $i\neq m\neq n$. 
The parameter $\alpha $ was introduced in order not to violate the sum rule
of correlation functions.
For the inhomogeneous case, the lattice translational invariance does not exist
and one needs to introduce $N$ such parameters for a lattice of $N$ sites.
However, simply replacing $\alpha$ by $\alpha_i$ leads to difficulty
in solving self-consistent equations of $G$.
Instead, we introduce $N$ parameters $\beta _{m}$ for each site $m$
according to the following relations:
\begin{eqnarray}
S_{m}^{+}&=&S_{m}^{+}[\beta_{m}-2(1-\beta_{m})S_{m}^{z}]
\nonumber\\
S_{m}^{-}&=&[\beta_{m}-2(1-\beta_{m})S_{m}^{z}]S_{m}^{-}
\end{eqnarray} 
for the spin$=\frac{1}{2}$ case.
The decoupling approximation for
the inhomogeneous systems is thus expressed as
\begin{eqnarray}
\langle \langle S_{m}^{+}S_{n}^{-}S_{i}^{z};~S_{j}^{z}\rangle \rangle
 &=& \langle \langle 
S_{m}^{+}[\beta_{m}-2(1-\beta_{m})S_{m}^{z}]\nonumber \\
&&[\beta_{n}-2(1-\beta_{n})S_{n}^{z}]S_{n}^{-};~S_{j}^{z}
\rangle \rangle \nonumber\\
&\rightarrow &\beta _{m}\langle S_{m}^{+}S_{n}^{-}\rangle \beta
_{n}\langle \langle S_{i}^{z};~S_{j}^{z}\rangle \rangle . 
\end{eqnarray}
Here we only keep terms of 
the lowest order (three operators) Green's function. 
On the right hand side of Eq. (7) two parameters 
$\beta_{m}$ and $\beta_{n}$ are attached
to the correlation function of the two corresponding spins 
on sites $m$ and $n$.
It is clear that these parameters are vertex corrections
of the spin-spin correlations.
According to the definition, we introduce one vertex
parameter for each site, and no other extra parameters.
In the uniform case we obtain $\beta ^{2}=\alpha$, and our decoupling 
approximation is the same as the KYDA (as shown in Eq. (5)). 

With the help of the above decoupling scheme, we 
obtain the equations of motion of the spin Green's function $G$
\begin{eqnarray}
\omega ^{2}G(i-j,\omega ) &=&\sum_{\eta }2J_{i,i++\eta }\{C(i,i+\eta
)(\delta _{i+\eta ,\ j}-\delta _{i,\ j})  \nonumber \\
&&+\langle \langle 2J_{i,i+\eta }(S_{i}^{z}-S_{i+\eta }^{z})+\Pi ;\
S_{j}^{z}\rangle \rangle \},
\end{eqnarray}
where 
\begin{eqnarray}
\Pi &=&\sum_{\eta \neq \eta ^{\prime }}\{2J_{i,i+\eta ^{\prime }}[\beta
_{i+\eta ^{\prime }}\beta _{i+\eta }C(i+\eta ,i+\eta ^{\prime
})S_{i}^{z}\nonumber \\
&&\hspace{0.5in}-\beta _{i}\beta _{i+\eta }C(i+\eta ,i)S_{i+\eta ^{\prime }}^{z}]
\nonumber \\
&&\quad \;+2J_{i+\eta -\eta ^{\prime },i+\eta }[\beta _{i+\eta }\beta
_{i}C(i+\eta ,i)S_{i+\eta -\eta ^{\prime }}^{z}
\nonumber \\
&&\hspace{0.5in}-\beta _{i+\eta -\eta
^{\prime }}\beta _{i}C(i+\eta -\eta ^{\prime },i)S_{i+\eta }^{z})]\} ~,
\end{eqnarray}
here the relation $C(i,j)=\langle S_{i}^{+}S_{j}^{-}\rangle
=2\langle S_{i}^{z}S_{j}^{z}\rangle $ has been used.

Since there exists no translational invariance, we can only solve
this set of equations of Green's function in real space. For a finite 
lattice, the Green's function $G$ can be
expressed in a matrix form $\widetilde{G}$, 
and Eq. (8) can be rewritten as
\begin{equation}
\omega^{2}\widetilde{G}-\widetilde{h}\ \widetilde{G}=\widetilde{C},
\end{equation}
where matrices $\widetilde{h}\ $ and $\widetilde{C}$ are made of the
nearest-neighbor, the second nearest-neighbor, and the third nearest-neighbor
spin-spin correlation functions.
We solve self-consistent equation (10) to determine the spin-spin correlation
functions and the vertex correction parameters $\beta _{i}$ for each site
so to calculate thermodynamical quantities at any temperature.
\vskip -1.5cm
\begin{figure}[ht]
\epsfxsize=3.0 in\centerline{\epsffile{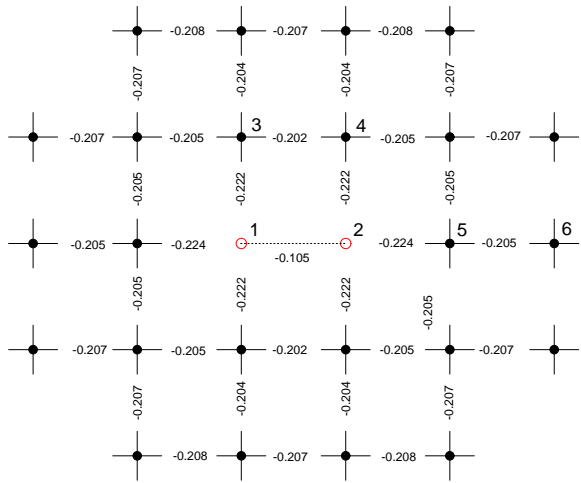}}
\caption{The NN correlation functions of the 8$\times 8$ lattice
with one broken bond at temperature $T=0.01(2J)$. The dotted line
between site 1 and 2 (open circles) represents the broken bond.}
\end{figure}

\section{Numerical Results}

We have performed  numerical calculation on the 6$\times $6, 
8$\times $8, and 10$\times $10 lattices with periodic boundary 
conditions. For the convenience of comparison, we first study 
the 6$\times $6, 8$\times $8, and 10$\times 10$ homogeneous 
Heisenberg antiferromagnet.
Our results of these finite lattices are quite close to the
results of an infinite lattice \cite{Shimahara}, since the interaction is
short-ranged. For the ground
state we obtained that the NN correlation function $C_{1}$ is 
equal to -0.2080 for 6$\times $6 lattice, which is quite 
close to the corresponding result $-0.2067$ for the infinite 
lattice \cite{Shimahara}. 
We also studied the staggered structure factor $S(Q)$ ($Q=(\pi ,\pi )$ is the
AF momentum), defined by,
\begin{equation} 
S(Q)=\frac{1}{N^{2}}\sum_{i_{x},i_{y}}\langle 
[(-1)^{i_{x}+i_{y}}S^{z}(i_{x},i_{y})]^{2}\rangle  ~,
\end{equation}
and found that its leading finite-size dependence is of order $1/L$,
which agrees with the scaling law proposed by other theories \cite{Huse,Liu}. 
The extrapolated estimate of the staggered magnetization for an 
infinite lattice is $m=\lim_{L\rightarrow\infty}\sqrt{S(Q)}=0.23$, 
which grossly agrees with the results of the other theories \cite{Huse,Liu}.
Even though we are limited to finite size clusters, our calculation can give
reasonable estimates for the infinite system
for problems we are interested in.

Let us study the one broken bond case. All configurations of 
one broken bonds are equivalent within the periodic boundary 
conditions. Our numerical results of the NN correlation 
functions near the broken bond for the 8$\times $8 lattice at 
temperature $T=0.01(2J)$ 
are shown in Fig. 1. Our results show that the broken bond enhance
correlation between spins close to it, which means that the
quantum fluctuation close to the broken bond is reduced.
The biggest NN correlation function is $C(2, 5)=-0.224$,
which is about $8\%$ lower than  $C_{1}=-0.207$ of the uniform case. 
At zero temperature, Lee and Schlottmann \cite{Schlottmann} 
have studied this model by using the LSW theory.
They showed that the quantum fluctuation is reduced in the neighborhood of 
the impurity link and the local magnetic moment is enhanced. 
Our results agree with the results of the LSW theory. 

The scaling effect of the NN correlation functions close to the 
broken bond is shown in Table I. We found that the results 
of the $8\times 8$ lattice are very close to that of the 
10$\times $10 lattice. We also found that the NN correlation 
functions in the next neighborhood of the broken bond are reduced, 
that is, the quantum fluctuation gets enhanced as the distance to 
the broken bond increases. 

\vspace{0.6cm}
\small
TABLE I. The scaling effect of the NN correlation functions 
of some special NN sites (labeled as in Fig. 1). 
The NN correlation function of uniform case is $C_{1}=-0.207$.   
\normalsize

\vspace{0.3cm}
\begin{tabular}{||l|c|c|c|r||}\hline\hline
$N\times N$  &  $C(1, 2)$ & $C(3, 4)$ & $C(2, 5)$ & $C(5, 6)$ \\ \hline
6$\times$ 6  &  -0.114    & -0.200    & -0.226    & -0.204    \\
8$\times$ 8  &  -0.105    & -0.202    & -0.224    & -0.205    \\
10$\times$ 10 & -0.101    & -0.203    & -0.224    & -0.205    \\ \hline
\end{tabular}
\vspace{0.7cm}

The energy cost for removing one bond from the homogeneous lattice
is defined as
\begin{equation}
\Delta_{1}=E^{1}_{0}-E^{0}_{0},
\end{equation} 
where $E^{0}_{0}$ is the ground state energy of the uniform case, and 
$E^{1}_{0}$ is the ground state energy of the one broken bond case. For the
8$\times $8 lattice, we obtained that $\Delta_{1}=0.10(2J)$, which is
smaller than the energy per bond $\frac{E^{0}_{0}}{2N}=-0.312(2J)$ of the
8$\times $8 homogeneous Heisenberg model. This different is due 
to the fact that the magnetic system has lowered its energy considerably 
by readjustment of its spin correlations.

\begin{figure}[ht]
\epsfxsize=2.2 in\centerline{\epsffile{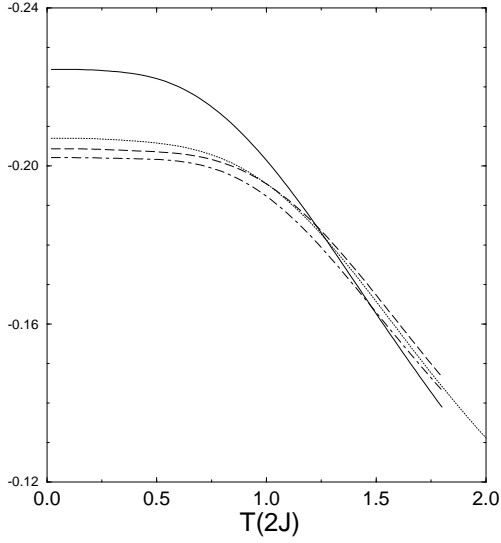}}
\caption{
The temperature dependent of the NN correlation functions 
$C(2, 5)$ (solid line), $C(5, 6)$ (dashed line) and 
$C(3, 4)$ (dash-dotted line) of 8$\times $8
lattice. The dotted line is the result of the homogeneous model.} 
\end{figure}
 
The temperature dependent of $C(2, 5)$, $C(5, 6)$ and $C(3, 4)$ 
are shown in Fig. 2. For the convenience of comparison, the 
temperature dependent of NN correlation 
\begin{figure}[ht]
\epsfxsize=3.0 in\centerline{\epsffile{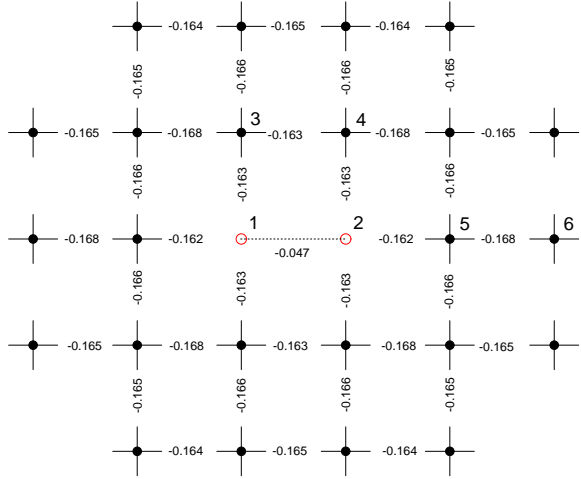}}
\caption{The NN correlation functions of 8$\times 8$ lattice
with one broken bond at temperature $T=1.5(2J)$. The dotted line 
between site 1 and 2 (open circles) represents the broken bond.} 
\end{figure}
\noindent function $C_{1}$  
of the homogeneous Heisenberg model is also plotted (dotted line).
In the low temperature region, the correlation function $C(2, 5)$ 
is larger than  $C_{1}$. When the temperature increases,  
$C(2, 5)$ drops very quickly and becomes smaller than  
$C_{1}$ as $T>1.25(2J)$. 
For the 8$\times $8 lattice, the NN correlation function near the 
broken bond at temperature $T=1.5(2J)$ are shown in Fig. 3. 
The biggest NN correlation function is $C(5,6)=-0.168$, which is 
only about $1\%$ lower than the NN correlation function of 
the homogeneous case ($C_{1}=-0.166$). It is obvious that the effect 
of broken bond on the nearby NN correlation functions in the high 
temperature region is weeker than that in the low temperature region. 

\begin{figure}[ht]
\epsfxsize=2.0 in\centerline{\epsffile{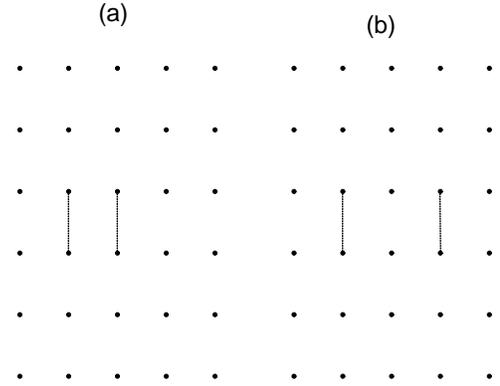}}
\caption{
Two different configurations of the two broken bonds case. 4(a) the 
two broken bonds are parallel and adjacent; 4(b) the two broken bonds 
are still parallel but have a distance of two lattice spacing.} 
\end{figure}

For the two broken bonds case, we mainly consider 
two configurations as shown in Fig. 4(a) and 4(b). 
In Fig. 4(a), the two broken bonds are parallel and adjacent. 
While in Fig. 4(b) the 
two broken bonds are still parallel but the distance between these 
two bonds increases to two lattice spacing. 
We obtained that the energy cost of constructing the configuration 
4(a) from the homogeneous lattice, e.g., removing two adjacent AF bonds, 
is $\Delta_{2a}=0.31(2J)$, which is higher than the energy of 
introducing two isolated broken bonds $2\Delta _{1}=0.20(2J)$. 
In addition, $\Delta _{2a}$ is quite smaller than the two bond energy 
$2\frac{E^{0}_{0}}{2N}=-0.624(2J)$ of the uniform case. 
The corresponding energy of the 
configuration 4(b) is $\Delta _{2b}=0.24(2J)$, which is 
more closer to $2\Delta _{1}$ than $\Delta_{2a}$. 
Thus configuration 4(b) has energy lower than that of 4(a).
Our results can be interpreted as that there exists effective
repulsive interactions between the two parallel broken bonds. 

\begin{figure}[ht]
\epsfxsize=2.8 in\centerline{\epsffile{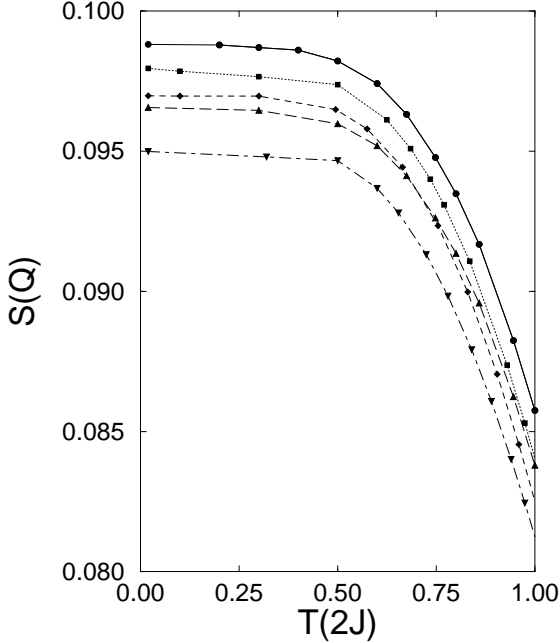}}
\caption{
The spin staggered structure factor $S(Q)$ as a
function of temperature $T$ for the homogeneous lattice (circle),
the one broken bond case (square), 
the two parallel broken bonds as Fig. 4(a) (diamond)
and Fig. 4(b) (up triangle),
and the three parallel adjacent broken bonds case(inverse triangle).} 
\end{figure}

The spin staggered structure factor $S(Q)$ ($Q=(\pi,\pi))$ as
functions of temperature $T$ for the homogeneous lattice and some 
particular configurations of one, two and three 
broken bonds cases are plotted in Fig. 5, respectively. 
The configuration of three broken bonds shown in Fig. 5 is that
the three broken bonds are parallel and adjacent.
Our results showed that, although the AF correlation functions
close to the broken bonds are enhanced, 
the AF order of the whole system will be suppressed
as number of broken bonds increase.

\section{Summary}

In conclusion, we have extended the second-order Green's function theory
to the inhomogeneous Heisenberg model by improving the decoupling 
approximation introduced by Kondo and Yamaji. 
The Kondo and Yamaji's decoupling approximation 
can not be directly expanded to the inhomogeneous case. 
In this paper,
we introduced an improved decoupling approximation, which is in
accordance with the Kondo and Yamaji's decoupling approximation
in the homogeneous case. In our approximaton two parameters are 
attached to the correlation function of the two corresponding spins,
which are vertex correction of spin-spin correlation
and they have to be introduced in 
order not to violate the sum rule of correlation functions. 
We have tried the one parameter scheme (Eq. (5)) and met difficulties
in getting the self-consistent equation (10) converge.
In our decoupling approximation, there are $N$ vertex correction parameters
for a lattice of $N$ sites and no other extra parameter was introduced
so we did not add one more parameter than the one parameter scheme.
Moreover, our decoupling approximation reduces to the Kondo and Yamaji's in
the homogeneous case.

We apply this theory to study the 2D Heisenberg antiferromagnet 
with broken bonds for all temperatures.
At low temperatures, our numerical results 
showed that the AF nearest-neighbor correlation functions close 
to the broken bonds are enhanced. That is, the quantum 
fluctuation is reduced in the neighborhood of the broken bond 
at low temperatures. Our results are in agreement with the 
results of other theories. As the distance to the broken 
bond increases, the NN correlation functions decreases.
By contrast, at high temperatures our results showed that the quantum
fluctuation close to the broken bond is enhanced.  
For the two broken bonds case, we found that there exists repulsive 
interaction between the two parallel neighbor bonds.

\section*{Acknowledgements}

We thank Prof. Shiping Feng, Prof. Qingqi Zheng and Dr. Liangjian Zou
for helpful discussions.
This work was supported in part by the Earmarked Grant for Research from the
Research Grants Council (RGC) of the Hong Kong Government
under projects CUHK 311/96P-2160068 and 4190/97P-2160089.

\end{document}